# Slowly evolving early universe and a phenomenological model for time-dependent fundamental constants and the leptonic masses

Dara Faroughy

*Department of Sciences: Pima College West Campus, Tucson Az USA.*

Contact email:  darifaro@yahoo.com

A phenomenological model with an extreme accuracy is proposed for the cosmic time variation of the primordial fundamental constants (e, h, G and c) and the leptonic masses. The model is purely exploratory in that at the very early times the light speed is purposely modeled to be negligibly small, indicating a very slowly expanding universe around t=0. The impact of this idea in cosmology and its modeling is overwhelming.

## Introduction

Undergraduate students learn early on during their study that physics at its fundamental level is characterized by a small number of dimensionful fundamental constants, e, h, c and G. Radiation theory exist because accelerating charged light sources have charges that are themselves expressed in unit of the electron charge e≠0. Quantum physics, describing nature at the microscopic level, exist because the Planck constant h, though very small, is not zero; special relativity and field theories exist because the ultimate speed of causal physics, the light speed c, is not infinity. And, finally classical gravity and its relativistic version (General Relativity) exist because the Newton's constant G is not zero. There are other coupling constants at the microscopic level (like in QCD) and more constants at the macroscopic level (e.g., the Boltzmann constant of thermodynamics, the Hubble, and the cosmological constants of cosmology, and so forth) that we omit discussing over here.

What we know is that most local laws of physics, as expressed in form of differential equations, involve, in various ways, the fundamental constants that are introduced by hand. Moreover, we also know that these local laws (usually derived from appropriate Lagrangian densities) are not generally invariant under an arbitrary global scaling (or local gauging) of the fundamental constants they include. (E.g., the Schrodinger, Klein-Gordon, Dirac, Rarita-Schwinger,… equations and the basic Poincare'algebra of relativistic quantum field theories, are not invariant under rescaling h→γh, or c→γc, and the spacetime Ricci tensor and the scalar curvature of general relativity are not invariant under G→γG, etc.). And obviously this means various predictions that we make in terms of physical observables are generally affected by what numerical values the constants may take.

But now comes the yet unsolved mystery: what determines these fundamental constants in the first place, abundantly quoted in most physics textbooks? This is a great question because for one, the local laws of nature do not generally determine the numerical values of the constants. And for another, there are no satisfactory theories at disposal that can yield the numerical values of these constants from first principles, along with predicting mass values of some fundamental particles comprising the Standard Model (SM), like the electron mass, by direct calculations.  The fermionic (i.e., leptons and quarks) and the bosonic masses (like the $W^{\pm}$ and $Z^o$) are predicted through the electroweak spontaneous symmetry breaking mechanism needing the Higgs mass as an input, along with few other experimentally determined input parameters, such as the Weinberg angle. The predicted parametric masses in the SU(3)xSU(2)xU(1) SM gauge theory are then compared to the observed masses and in this way useful information is extracted for the electroweak many parameters and the Higgs mass. Many "group theoretic" extensions of the SM gauge theory are also available in form of, e.g., Grand Unified Theories (GUT), SupersymmetricSM, SupergravitySM theories, and so on, also making their own specific predictions for the particle masses and ratios through generalized form of the Higgs mechanism. However, such theories have nothing to say about the values of the fundamental constants, which are again introduced by hand in the grand Lagrangians. Normally, if the constants were to vary in spacetime, then they would have to be treated as fields and expectedly that would complicate further whatever Lagrangians/actions may be envisaged for a given situation.

So, it seems that in the absence of any comprehensive theory regarding fundamental constants we are left with only unanswered questions and here are few that comes to mind. Are the measured constant values somewhat accidental, or say antropically tailored (note: in a multi universe scenario, e.g., the constants of physics can be distinct in diverse universes yielding distinct particle behavior and different atomic physics and chemistry altogether, and if so then what becomes of life!)? Do they have their current experimental values because they are somehow related to the microscopic laws of physics, or some topology or perhaps to our evolving universe, which happens to be a dozen billion years old? For the latter case then does this means that the constants had different values long time ago (we note in passing that string theory, e.g., demands variation of the dimensionful physical parameters because of the presence of the dilaton scalar field in the theory) when the universe was in its infancy and (presumably) very rapidly inflating and undergoing a series of phase transitions (in fact we expect a GUT phase transition when the vacuum energy density is about $10^{91}$ J/m$^3$, an electroweak phase transition when the energy density is about $10^{11}$ J/m$^3$, etc)? Expectedly nobody knows the answer! Consequently, any present attempt to say "model" these constants is forcibly phenomenological in essence, and that's the story for the time being! In this text we like to use certain dimensionless constants of mathematics (like π~3.14 and the e~2.7) to model the constants with seemingly some topological (yet unknown to us) reasons behind their existence in an early universe. After all who can say a spin ½ electron or a spin 1 photon, all expressed in unit of ħ, existed at t=0 if nature didn't know what the number π meant at t=0 (recall ħ=h/2π)! A spacetime evolving π has been of interest to this author for some years **[1]** but its complex discussion is outside the scope of the present short paper.

Heuristically, an early phenomenology for the variation of, e.g., the Newton constant G with the cosmic time t was proposed long ago by Dirac in 1933, where G~1/t (thus making G small for an old universe). Decades later Brans-Dicke **[2]**, inspired by the Mach principle, extended the Dirac idea to GR and proposed a spacetime variable model for G (while keeping other constants constant), nowadays also known as the scalar-tensor model of gravity, as an alternative to GR that is still with us today. The BD action in the Jordan frame involving the physical scalar field Φ=Go/G (where Go is the bare gravitational constant) and matter field Lagrangian $L_M$ is:

$$S = \frac{1}{16\pi} \int d^4x \sqrt{-g} \left(\phi R - \omega \frac{\partial_a \phi \partial^a \phi}{\phi} + \mathcal{L}_M\right)$$

Where R is the scalar curvature, $\omega$ is the dimensionless coupling and g is the determinant of the metric. Extremizing the BD action yields the field equations in below, where T is the trace of the stress-energy tensor $T_{ab}$ and $G_{ab}=R_{ab}-\frac{1}{2}Rg_{ab}$:

$$G_{ab} = \frac{8\pi}{\phi} T_{ab} + \frac{\omega}{\phi^2}\left(\partial_a\phi\partial_b\phi - \frac{1}{2}g_{ab}\partial_c\phi\partial^c\phi\right) + \frac{1}{\phi}(\nabla_a\nabla_b\phi - g_{ab}\Box\phi)$$

$$\Box\phi = \frac{8\pi}{3+2\omega}T$$

The current experimental limit on time variation of G, e.g., is about $\dot{G}/G < -10^{-12}$ yr$^{-1}$. And the observational limit on the BD dimensionless coupling constant $\omega$ may vary in the range of few dozens to few thousands, depending on what (generally) cosmic phenomenon is observed (we note in BD theory $\dot{G}/G \sim -H/\omega$, where H is the Hubble constant). In case of a flat universe with RW metric (and k=0) the BD theory predicts a time variation for G as $G \sim t^{-2/(4+3\omega)}$. The BD model is only one model among others that yield variable G from first principles per se.

Our model, presented shortly, is purely phenomenological in nature and as such its underline principle is still lacking. Yet, the excellent compatibility with the observed values of the fundamental constants and the electron mass, given the startling mathematical simplicity of the model, is perhaps an indication that some truth may lie behind the model. There is obviously much more to the very long, multifaceted and intriguing story of variable constants (and not to omit arguments on how life, as we know it today, could have been different or perhaps nonexistent if these constants had different present values). The grand unified theories (the best being the supersymmetric GUT) unify the three nongravitational coupling constants at some high-energy scale. So it is reasonable to expect that any change of one or more of the fundamental constants (e.g., affecting the fine structure constant $\alpha=e^2/\hbar c \sim 1/137$) would also impart changes in the QCD and the weak coupling constants. But the goal of this short note prevents us expounding further into the general topic of variable fundamental constants so that we can present our rather amusing model of the (cosmic) time variation of h, c, e and G, yielding the observed present values of these constants, as well as an effective time variable Planck length which is supposedly relevant to quantum gravity!

The quoted present values of the constants and the observed electron mass in the cgs units are as follows: c=2.99792458x10$^{10}$cm/s, $\hbar$=1.05457168x10$^{-27}$ergxs, e=4.80320441x10$^{-10}$esu, $m_e$=9.1093826x10$^{-28}$g, $\alpha^{-1}$=137.03599911, $G_N$=6.67425211x10$^{-8}$, and also $\pi$=3.1415926535898, ex=2.71828182445904. We define k= $\pi$/2ex=0.577863674895325 and fix the present age of the universe as $t_u$=12.8891282 billion years=4.0674995208432x10$^{17}$ s, which is quite reasonable. (Interested readers can play around slightly with the value of $t_u$ and use the formulas we shall present shortly to see what ensues.) We also set $t_0$=1 light year=3.15576x10$^{16}$ s. Finally, we introduce a constant $\beta=\alpha^\sigma$ where $\sigma=2(1+0.94444\pi^2\alpha)$ to be used for evaluating the electron time variable mass.

## The Model

I shall assume throughout that the fine structure constant α, dominating atomic physics and the electromagnetic interaction, remains independent of the cosmic time (and that despite some experimental findings indicating possible very minute changes of α) so to not affect, e.g., the QCD scale factor Λ, or the weak coupling. We propose the following simplistic expression for **α=(9/16)(5!)$^{-0.25}$π$^{-b}$,** where b=ex+π$^{-3.01458214705}$, yielding α$^{-1}$=137.03599911087!

Moreover, I shall take a different, yet perhaps significant, attitude here regarding the very early universe in that I shall assume the universe begin very slowly and not ultra fast as in the inflationary and the more updated traditional scenarios. To implement this idea we shall let the light speed c(t) vanish at t=0, but this, as we shall see soon, does not mean very early particles are deprived of, e.g., rest energies, e.g., we shall see that the rest energy of an electron at t=0 is infinity even though c(0) vanishes! A slowly evolving early universe (expansion wise) obviously presents a new challenge to cosmologists in terms of modeling. What we shall find in below, e.g., is a universe that near t=0 is "numerically" very much dominated by very massive rest mass particles, strong h (thus QM), strong G (thus gravity), strong charge e (thus E&M, but note α is always kept small), strong elementary magnetic monopole charge g (note: by virtue of the Dirac quantization rule g$_o$=½ħc/e and by insisting α to be a constant we find g$_o$(t)/e(t)~68.518, thus a constant at all times), and all this while the light speed c is very small near t=0. Thus the challenge!

In short, here are our phenomenological expressions for the time dependent constants and the electron mass (as usual all units are cgs units):

$c(t)=2t^k$
$G(t)=t^{k-1}[\pi+(1+25\alpha)/(\pi.ex)]^{1/2}$
$h(t)=4\pi^2 t^{-(k+1)}[1+\pi.\alpha+(1.010449.\alpha.ex)^2]$
$e(t)=(\alpha.h(t).c(t)/2\pi)^{1/2}$
$m_e(t)=e(t)[\pi.h(t)/G(t)c(t)^5]^{¼(1-\beta)}$

The above expressions yield the following "present" values where t=t$_u$:

$C(t_u)=2.99792458 \times 10^{10}$
$G(t_u)=6.67427008 \times 10^{-8}$
$e(t_u)=4.8032044 \times 10^{-10}$
$h(t_u)=6.6260693 \times 10^{-27}$
$m_e(t_u)=9.1093826 \times 10^{-28.}$

As seen, the predicted numbers are by design exactly equal to the experimental values at t=t$_u$. Also of importance is the finding G$^.$/G~-3.275x10$^{-11}$/year.

The graph in below shows the time variation (expressed in billion years) of the ratios of the fundamental constants and the electron mass to their present values. However, the numerical values near t~0 are substantially higher (or lower in case of c(t)) than is shown in the graph. As also seen there are numerical similarities (though only above t=1 billion years) between the

Planck ratio and the electron mass ratio cosmic time variations on one hand and between the gravity ratio and the charge ratio on the other hand, certainly having some theoretical significance! The electron Compton length $h(t)/m_e(t)c(t)$ time changes are interesting. Indeed the latter is 8.6270725 cm at cosmic time t=1 s, while at $t=10^{-43}$ s it is $4.45742 \times 10^{28}$ cm (!) and at the present ($t=t_u$) it is $3.861597648 \times 10^{-11}$ cm. One billion years after the "slow" bang it was $\sim 2 \times 10^{-10}$ cm while for t=100 billion years it is reduced to $1.031125 \times 10^{-11}$ cm. Of interest to quantum gravity and the string models is the so-called time variable Planck length $L_p(t) = (\hbar(t)G(t)/c^3(t))^{1/2}$. At $t=10^{-43}$ s $L_p = 2.258 \times 10^{80}$ cm (!) while at t=1 s is 1.20648647 cm and today ($t=t_u$) it is $1.162512966 \times 10^{-33}$ cm. Another significant finding is the rest energy of the electron shown by $E_e(t) = m_e(t)c^2(t)$. Let me quote in below few numerical results all in unit of erg: $E_e(10^{-70}) = 1.145 \times 10^{25}$, $E_e(1) = 1.4906$, $E_e(10^{-2}t_0) = 1.04432 \times 10^{-5}$, $E_e(t_0) = 2.0315744 \times 10^{-6}$ and finally $E_e(t_u) = 8.1871048 \times 10^{-7}$ erg. Thus, the electron (and maybe more importantly the proton) looses rest energy to the environment as the universe ages!

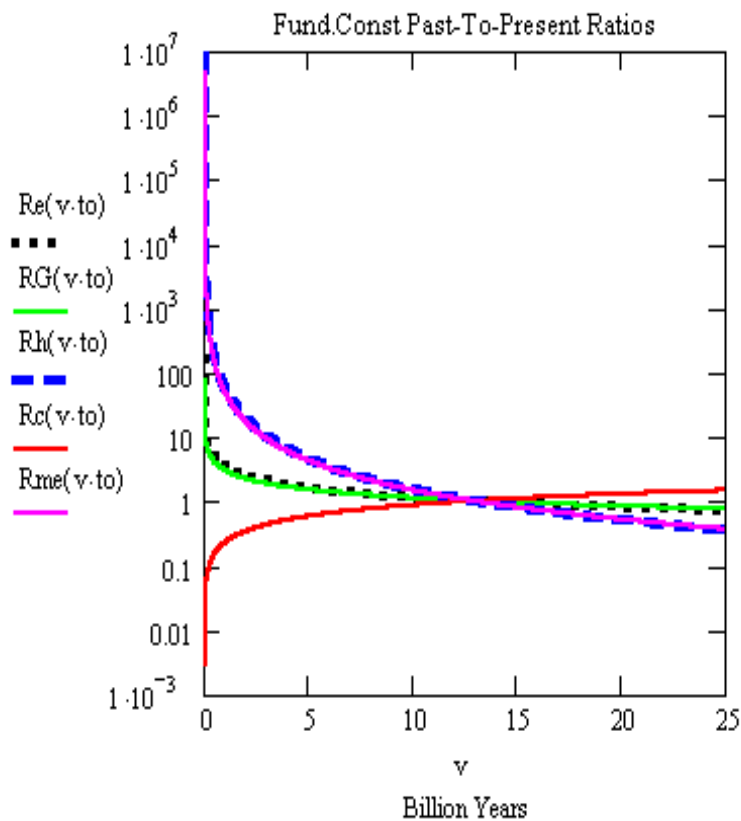

We are also capable of computing other quantities, like $e(t)h(t)/m_e(t)c(t)$ which is a quantity proportional to the Dirac electron magnetic moment, or the hydrogen binding energy, etc., as functions of t. For the former quantity $e(t)h(t)/m_e(t)c(t)$ we, e.g., we find the following numerical values: $3.589 \times 10^{35}$, 16.60497, $2.1733 \times 10^{-18}$, and $1.16540639 \times 10^{-19}$ for $t=10^{-30}$, 1, $t_0$ and $t_u$ respectively. An interesting topic is how the proton mass scales compared to the time variable electron mass. By assuming a constant ratio for $m_p(t)/m_e(t)$ (~1836) for all time we may get some ideas about the effective mass of the confined quarks as a function of the cosmic time. Currently we believe more than 90% of the proton mass resides in the gluonic field energy and have some ideas on the u and the d quark mass contributions to $m_p$. Assuming the same proportionality of quark mass contribution to $m_p$ holds during the ages we can relate the "present" u and the d quark masses to their values at any other times (details omitted). Obviously armed with the above formulas we can do much more than what we've presented so far. An intriguing thing we can do is to explore some of Dirac's original expression he proposed in 1938 in an attempt to connect quantum physics to the cosmological parameters. One such a relation, e.g., is the Hubble constant that Dirac proposed as $H=Gm_e m_p c/h^2$. Another relation may be inventing a time dependent energy density using G, h and c and a mass scale m, like $G(t)m^6(t)c^4(t)/h^4(t)$ and see how it evolves with time. Although this can be done easily one should not expect the same type of interpretation, as is commonly carried out, for a very slowly evolving early universe model that we are proposing over here!

*Leptonic masses:* In proposing a phenomenological formula for the "present" value leptonic masses we shall use two inputs; (1) a leptonic mass scale $M=6.1237498 \times 10^9$ ev and (2) a parameter $s=0.23109835$ which we identify with $\sin^2(\theta_W)$, where $\theta_W$ is the Weinberg angle of the electroweak gauge theory. The proposed formula in below is sensitive to the integer n. The leptonic mass scale M (which may or may not correspond to any physical heavy lepton) corresponds to n=0, the tau lepton goes with n=1, the muon goes with n=2 and the electron with n=3. Beyond n≥4 we speculate the very light leptons are the neutrinos. E.g., we may suggest n=4 correspond to the tau neutrino, n=5 goes with the muon neutrino and n=6 goes with the electron neutrino (for n=7 the mass is exceedingly small~$5.7 \times 10^{-12}$ ev!). The negative integers –1, -2… are also interesting to some extent for the neutrino purposes, especially for n=-4, -5 and –6. Our "exponential" mass formula is: $\mathbf{M_n = M \cdot \exp[n^2(-1+s^2) - s \cdot n(4s - \frac{1}{3}(-1)^n)]}$. The predictions of the latter formula in ev for n=1, 2, 3, 4, 5 and 6 are as follows:

$M_1 = 1.776992797 \times 10^9$
$M_2 = 1.056751406 \times 10^8$
$M_3 = 5.109989253 \times 10^5$
$M_4 = 9.378327463 \times 10^2$
$M_5 = 7.557035602 \times 10^{-2}$
$M_6 = 4.280308632 \times 10^{-6}$.

The experimental values for the tau, muon and the electron masses are 1776.99, 105.658369 and .51099892 Mev respectively. As seen the predictions are in excellent agreement with the data.

In case the above tiny neutrino masses of the muonic and the electronic types are ruled out by the finalized experiments (as too small) then we may suggest using negative integers for the neutrino masses with n ≤-4. For example, we find the somewhat more reasonable neutrino mass values: $M_{-4} = 2.797036306 \times 10^3$ ev, $M_{-5} = 1.38250145$ ev and $M_{-6} = 2.204622 \times 10^{-5}$ ev for the tau, muon and the electron type neutrinos respectively. In either positive or negative integer cases as seen the electron neutrino is predicted to be of very small mass of around $10^{-5}$ ev!

Finally, it is clear that by using the above present value leptonic masses we can find their cosmic time variations by simply connecting with our earlier formulas (detailed omitted). For some recent general arguments on the possible meaning of particle masses in physics readers may want to consult **[3]**.